\documentclass[aps,prd,superscriptaddress,12pt,notitlepage]{revtex4}
\usepackage{graphics}
\usepackage{graphicx}
\usepackage[thinlines]{easytable}
\usepackage{amsmath,amssymb,mathrsfs}
\usepackage{caption}
\usepackage{epsf}
\usepackage{epsfig}
\usepackage[font=small,skip=0pt,justification=raggedright]{caption}
\usepackage{xcolor}

\newcommand{\insertplot}[5]{\begin{figure}
 \hfill\hbox to 0.05in{\vbox to #5in{\vfill
 \inputplot{#1}{#4}{#5}}\hfill}
 \hfill\vspace{-.1in}
 \caption{#2}\label{#3}
 \end{figure}}
 \newcommand{\inputplot}[3]{
 \special{ps: plotfile #1}
}
\newcounter{fig}

\newcommand{\bea}{\begin{eqnarray}}
\newcommand{\eea}{\end{eqnarray}}
\newcommand{\be}{\begin{equation}}
\newcommand{\ee}{\end{equation}}

\newcommand{\re}[1]{(\ref{#1})}

\newcommand{\CL}{\ensuremath{\mathcal{L}}}

\newcommand{\CA}{\ensuremath{\mathcal{A}}}

\newcommand{\pa}{\partial}

\usepackage{epsfig}
\usepackage{amsmath}
\usepackage{amsfonts}
\usepackage{graphicx}
\usepackage[german, english]{babel}
\usepackage{amssymb}
\usepackage{ifthen}

\usepackage{ulem}
\usepackage{color}

\begin{document}

\title{Magnetic Hopfions in the Faddeev-Skyrme-Maxwell model}

\author{A.Samoilenka$^{\star}$ and Ya. Shnir$^{\dagger   }$}
\affiliation{$^{\star}$Department of Theoretical Physics and Astrophysics,
Belarusian State University, Minsk 220004, Belarus\\
$^{\dagger}$BLTP, JINR, Dubna 141980, Moscow Region, Russia}

\begin{abstract}
We construct new solutions of the Faddeev-Skyrme-Maxwell model, which
represent Hopf solitons coupled to magnetic fluxes.
It turns out that coupling to the magnetic field allows for transmutations of the
solitons, however, the results depend both on the type of the vacuum boundary condition and
on the strength of the gauge coupling.
It is shown that the structure of the magnetic fluxes of a gauged Hopfion
is governed by the preimages of the points
$\phi_3=\pm 1$.

\end{abstract}
\maketitle

\section{Introduction}

Topological solitons appear as classical solutions in various non-linear models, they
have been intensively studied over last decades. These regular localized field
configurations with finite energy attracted a lot of attention, they emerge in variety of physical, chemical and
biological systems.

Interesting examples of stable topological solitons exist in the family of
Skyrme-type scalar theories, which can be considered as deformations of the non-linear
sigma model. It includes so-called baby Skyrmions in (2+1)-dimensional
$O(3)$ model \cite{BB,Leese:1989gj}, Skyrmions in the conventional
(3+1)-dimensional Skyrme model \cite{Skyrme:1961vq} and its modifications
\cite{Adam:2010fg,Adam:2010ds},  and the Hopfions in the Faddeev-Skyrme model
\cite{Faddeev-Hopf,Faddeev:1996zj}.
A unifying feature of all these models is that the have the same structure, the corresponding Lagrangians
always include the usual $\sigma$-model term, the Skyrme term, which is quartic in derivatives of the
field, and a potential term, which does not contain the derivatives.
According to the Derrick's theorem
\cite{Derrick} the potential is optional
in $3+1$ dimensions, however it is obligatory to stabilise the soliton solutions
of the planar baby-Skyrme model.

The solitons of the Faddeev-Skyrme model are somewhat special because their
topology is defined by the first Hopf map  ${S}^3 \mapsto {S}^2$ with
the related homotopy group $\pi_3(S^2) = \mathbb{Z}$. It corresponds to the topological charge,
which the linking number of loops on the compactified domain space $S^3$.

Notably, all the models of the Skyrme family on non-compact domain
do not saturate the topological bound.
In order to attain the bound which yields a relation between the static energy of the solitons and
their topological charges $Q$, one has to modify the model preserving its topological properties, for
example, truncate the Faddeev-Skyrme model \cite{Foster:2010zb},
or oppositely, extend the Skyrme model via coupling it to an infinite tower of vector mesons
\cite{Sutcliffe:2011ig}, or completely change the original theory to the form, which supports self-dual equations
\cite{Adam:2010fg,Adam:2010ds,Ferreira:2017yzy,Ferreira:2017bsr}.
Thus, the energy of interaction between the solitons is relatively large, they may
attract each other forming various multisoliton configurations, see e.g. \cite{Manton-book,Shnir-book}.

Intuitively, the Hopfions can be constructed by considering baby Skyrmions restricted to the plane,
which is orthogonal to the direction of the position curve of the string-like configuration \cite{Kobayashi:2013xoa}.
The topological charge of such a soliton corresponds to the product
of the winding number of the planar Skyrmions and the number of the twists of the entire configuration in the extra spatial direction.
Physically, solitons of that type can be considered as a vortex, which is bending  and twisting. The
identification of the end points of the vortex yields the loop, which can transform
itself into a knot to minimise its energy.

A peculiarity of the interaction potential, both in the case of Skyrmions and Hopfions, is that the asymptotic
decay of the fields, which defines the character of interaction, strongly depends to the explicit form of the
potential \cite{Adam:2010fg,Adam:2010ds,Battye:2004rw,Battye:2009ad,Salmi:2014hsa,Gudnason:2016cdo,Gudnason:2016tiz}.
Further, various symmetry breaking potentials were considered to construct
half-Skyrmions \cite{jaykka2010easy,kobayashi2013fractional,kobayashi2014vortex,Gudnason:2015nxa,Lukacs:2016mvy}
or fractional Hopfions \cite{Samoilenka:2017hmn}.

There is another possibility to make alterations to the structure of multisoliton solutions.
In the Faddeev-Skyrme model the vacuum boundary condition should be imposed in such a way that
all the points on the boundary are identified. It yields the compactification of the domain space
from $\mathbb{R}^3$ to $S^3$. Hence the Hopfions are invariant with respect to the global $SO(2)$ symmetry of the vacuum.
This allows us to construct the $U(1)$ gauged Faddeev-Skyrme-Maxwell theory by analogy with the extension of the
gauged planar baby Skyrme model \cite{Samoilenka:2015bsf,Gladikowski:1995sc,Adam:2017ahh}.
Clearly, electromagnetic interaction will strongly affect the usual pattern of
interaction in the system of Hopfions.

Unfortunately, the task of explicit construction of the solutions of the Faddeev-Skyrme-Maxwell theory has
been hampered by numerous technical obstacles. Since there is no analytical solutions of
the corresponding field equations, the minimizers of the corresponding energy functional can only be obtained numerically.
However, it is known that the Hopfions of lowest degrees $Q=1,2$
are axially symmetric \cite{Gladikowski:1996mb,Battye1998,Sutcliffe:2007ui}, thus in the paper \cite{Shnir:2014mfa}
the consideration was restricted to the case of the static axially symmetric gauged unlinked  Hopfions
${\cal A}_{1,1}$ and ${\cal A}_{2,1}$.
Assumption of axial symmetry simplifies the consideration significantly since the problem then can be reduced to the numerical
solution of system of coupled ordinary differential equations.
However, this symmetry is not a general property of general solutions of the Faddeev-Skyrme model supplemented by the Maxwell term,
thus this problem should be revisiting.

In this paper we investigate the structure of multisoliton solutions of full coupled Faddeev-Skyrme-Maxwell system.
Usually there is an ambiguity in the choice of the topological boundary conditions on the
scalar field, however, in the $U(1)$ gauged Faddeev-Skyrme model it becomes dependent on the
definition of the electromagnetic group. We consider two
choices of the vacuum boundary conditions $\vec \phi_\infty = (0,0,1)$ and
$\vec \phi_\infty = (1,0,0)$.
In both cases we perform full 3d numerical computations to find the corresponding magnetic Hopf solitons
in the sectors of degrees up to $Q=8$.
We study numerically the dependence of masses of the Hopfions and the corresponding magnetic fluxes
on the gauge coupling constant. We confirm, that in the strong coupling limit
the magnetic fluxes of the Hopfion become quantized in units of $2\pi$.

We found that in a general case the magnetic fluxes of gauged Hopfions are defined by the preimages of the vectors
$\vec \phi = (0,0,\pm 1)$, there is an intrinsic interplay between the topology of the Hopf map and the structure of
the magnetic field of the configuration.

The rest of the paper is structured as follows. In the next section we
briefly describe the Faddeev-Skyrme-Maxwell model.
In the Section 3, for the sake of completeness, we review the rational map approximation used as input in our numerical
simulations. Numerical results are presented in Section 4, where we
describe various magnetic Hopfion solutions. For
the sake of compactness, we restrict the analysis to the solitons with
topological charges up to eight, as a particular example we present a more detailed discussion of evolution of the $Q = 5$
Hopfions. Conclusions and remarks are
formulated in the last Section.

\section{$U(1)$ gauged Faddeev-Skyrme model}
We consider the Faddeev-Skyrme theory coupled to the Abelian gauge field in $(3+1)$ dimensions.
The model is defined by the rescaled Lagrangian
\be
\label{model}
\CL=\int d^3 x\left[-\frac{1}{4g^2} F_{\mu\nu}^2 + D_\mu \vec\phi\cdot D^\mu \vec\phi -
\frac{1}{2}\left(D_\mu \vec\phi\times D_\nu \vec\phi\right)^2\right] \, ,
\ee
where the real scalar triplet $\vec \phi = (\phi_1,\phi_2,\phi_3)$  is constrained to the surface of unit sphere,
$|\vec \phi \cdot \vec \phi| =1$, so the target space is the sphere $S^2$. Since the potential
term is optional, we do not consider it. However, the global $SO(3)$ symmetry will be broken as we impose
the topological vacuum boundary conditions, like $\vec \phi_\infty = (0,0,1)$,
which yield a one-point compactification of
the domain space $\mathbb{R}^3$ to $S^3$. Note that this common choice is not unique, below we will also
consider another case, $\vec \phi_\infty = (1,0,0)$.

Thus, the field of the Hopfion is a map
$\vec \phi:\mathbb{R}^3 \to S^2$ which belongs
to an equivalence class characterized by the homotopy group $\pi_3(S^2)=\mathbb{Z}$. Explicitly, the
Hopf invariant is defined non-locally as
\be
\label{charge}
Q=\frac{1}{16\pi^2}\int\limits_{\mathbb{R}^3}\varepsilon_{ijk}{\cal F}_{ij} {\cal A}_k
\ee
where ${\cal F}_{ij}=\vec \phi \cdot (\partial_i \vec \phi \times \partial_j \vec \phi )$
and one-form ${\cal A} = {\cal A}_k dx^k$ is
defined via ${\cal F}=d {\cal A}$, i.e the two-form ${\cal F}$ is closed, $d{\cal F}=0$.

The model \re{model} includes also the usual Maxwell term with the field strength tensor
$F_{\mu\nu}=\partial_\mu A_\nu -\partial_\nu A_\mu$. Note that
under a spacial rescaling ${\bf x} \to \lambda {\bf x}$, this term in the action
scales as $\lambda^{-1}$, i.e. it has the same scaling properties as the quartic in derivatives Skyrme term.
The flat metric is
$g_{\mu\nu}=diag(1,-1,-1,-1)$ and the coupling of the scalar triplet to the gauge field is
given by the covariant derivative \cite{Samoilenka:2015bsf,Shnir:2014mfa,Schroers:1995he,Gladikowski:1995sc}
\be
D_\mu\vec{\phi}=\pa_\mu\vec{\phi}+A_\mu\vec{\phi}\times\vec{n} \, ,
\label{gauging}
\ee
where the unit vector $\vec n = (0,0,1)$ defines the direction of the electromagnetic subgroup. Explicitly,
\be
\label{covder}
D_\mu\phi_\perp=\pa_\mu\phi_\perp-i A_\mu\phi_\perp, \ \ \  D_\mu\phi_3=\pa_\mu\phi_3
\ee
where $\phi_\perp = \phi_1 + i\phi_2 $ are planar components of the scalar field. Thus, the third component
$\phi_3$ remains decoupled from the gauge potential.
However, since the scalar triplet is restricted to the surface of the unit sphere, coupling
of the planar components $\vec\phi_\perp$ to the gauge sector still affects the component $\phi_3$ indirectly.

The Abelian gauge transformations act on the fields as
\be\label{gauge}
\phi_\perp\rightarrow e^{i \alpha}\phi_\perp, \ \ \ A_\mu\rightarrow A_\mu+\pa_\mu\alpha
\ee
thus, we can make use of this symmetry to set $A_0=0$. Further, restricting our analysis to static
configurations, we consider purely  magnetic field
$\vec B=(-\pa_3 A_2\, , \pa_3 A_1\, , \pa_1 A_2-\pa_2 A_1)$.

The static energy functional of the  model \re{model} is
\be\label{en}
E= \int d^3 x\left[ \frac{1}{2g^2}\vec{B}^2 +D_{i}\vec{\phi}\cdot D_{i}\vec{\phi}
+\frac{1}{2}\left(D_i \vec{\phi}\times D_j \vec{\phi}\right)^2 \right] \, .
\ee
Here we are using normalized units of energy, rescaling it as $E \to E/(32\pi^2\sqrt 2)$.
The Hopfions correspond to the stationary points of this  functional.
Note that the condition of finiteness of energy implies that
$
D_i\phi_\perp = \pa_i\phi_\perp - i A_i \phi_\perp \underset{r\to\infty}{\longrightarrow} 0\,
$ as $r\to \infty$. In other words, on the spacial asymptotic
the field of the gauged Hopfion must lie in an orbit of the gauge group, it is not necessarily a constant there.

The complete set of the field equations, which follows from the variation of the action of the
model \re{model}, is
\be
\begin{split}
\label{ELeqs}
    &D_\mu\vec J^\mu=0 \, ;\\
    &\pa_\mu F^{\mu\nu}-2g^2\vec{n}\cdot\vec{J^\nu}=0 \, .
\end{split}
\ee
Here the scalar current is
\be
\vec J^\mu=\vec{\phi}\times D^\mu\vec{\phi}-D_\nu\vec{\phi}(\vec{\phi}\cdot D^\mu\vec{\phi}\times D^\nu\vec{\phi}) \, ,
\label{current}
\ee
and a source in the corresponding Abelian Maxwell equations is $j_\mu = \vec n\cdot \vec J_\mu$. This system is
similar to the corresponding equations of the planar Skyrme-Maxwell theory
\cite{Schroers:1995he,Gladikowski:1995sc,Adam:2017ahh}, however, the topological properties of the fields are different.

Unlike other solitons, the location of the Hopfions does not correspond to the maximum of the topological charge density, the
Hopfions are extended string-like configurations in 3 dimensional space. The maxima of the energy density distribution can
be identified as the curve of positions of the preimage of the
point  $\vec\phi_0 = (0,0,-1)$, which is antipodal to the vacuum \cite{Sutcliffe:2007ui}.
This curve is usually referred to as the position curve \cite{Sutcliffe:2007ui}. In the gauged Faddeev-Skyrme model this curve
has also another meaning.

Note that we can make use of the trigonometrical
parametrization of the scalar field
\be
\vec \phi = (\sin \psi \cos \sigma\, ,
             \sin \psi \sin \sigma\, ,
             \cos \psi) \, ,
\label{trigonometric}
\ee
where two functions $\psi(x,y,z)$, $\sigma(x,y,z)$
satisfy the boundary conditions on the Hopfion configuration in a given topological sector.
Although this parametrization is not the most convenient  from the point of view of numerical simulations
\cite{Shnir:2009ct}, it automatically takes into account restriction of the scalar field to $S^2$.
This in particular, allows for a more transparent understanding of many peculiarities of the gauged
Faddeed-Skyrme model.

In a simple case of the axially symmetric gauged Hopfions ${\cal A}_{1,1}$ and ${\cal A}_{2,1}$
\cite{Gladikowski:1996mb}, the function $\sigma$  can be explicitly written in
spherical coordinates  $(r,\theta,\varphi)$ as $\sigma = n\varphi - m G(r,\theta)$, here
two winding numbers $n,m \in \mathbb{Z}$ correspond to the planar winding and the
twisting of the configuration, respectively. The phase function of the axially symmetric configuration
$G(r,\theta)$ increases by $2\pi$ after one revolution around the core, thus the Hopf number of the soliton
is just a product of two windings, $Q=mn$ and the axially
symmetric configuration of the type ${\cal A}_{m,n}$ can be thought of as composed from
planar baby Skyrmion of charge $n$ twisted $m$ times along the circle \cite{Gladikowski:1996mb}.

In a general case, by analogy with the similar situation in the
gauged planar Skyrme model \cite{Samoilenka:2017fwj}, the Abelian current can be written as
\be
j_i= (\pa_i\sigma-A_i)\left[ 1-\phi_3^2 +\pa_j\phi_3^2\right]  -\pa_i\phi_3\pa_j\phi_3(\pa_j\sigma-A_j)
\label{electocurrent}
\ee

We can assume that the gauge potential $A_i$ slowly varies in space. Then,
from the second equation in \re{ELeqs}, we can see that,
in the limit of infinitely large gauge coupling
$g\to \infty$ this equation is satisfied only if the Abelian current \re{electocurrent} becomes zero.
Evidently, if $\phi_3 \neq \pm 1$, the current is vanishing when $A_i=\pa_i\sigma$, i.e. the magnetic potential
becomes a pure gauge everywhere in 3d space apart the curves ${\cal C}_\pm = \phi^{-1}(0,0,\pm 1)$.

Considering
the magnetic flux through the area, transverse to the direction of the $\phi_3$, we can see that
\be
\Phi=\int B\, d^2x=\oint_\Gamma \vec{A}\cdot\vec{dl}=
\oint_\Gamma \nabla\sigma\cdot\vec{dl}= 2\pi n \, ,
\ee
where $\Gamma$ is a closed contour encircling the points, where $\vec\phi = (0,0,\pm 1)$.

The consideration above explains the effective quantization of the magnetic fluxes of the gauged
axially symmetric Hopfions  of degrees $Q=1,2$ in the strong coupling limit \cite{Shnir:2014mfa}.
It was observed that  the configuration is associated with
two magnetic fluxes, one of which represents a circular vortex, and the second one is orthogonal to the position
curve\footnote{More precisely, the direction of the second flux is given by the vector $\vec n$, which we
introduced in the definition of the covariant derivative \re{gauging}.}. In the strong coupling limit the
former flux is quantized in units
of the winding number $n$ while the latter flux is quantized in units of $m$.

Indeed, the position curve of the Hopfion is defines as the
preimage of the point  $\phi = (0,0,-1)$ on the target space. On the other hand, for the axially symmetric
Hopfions the component $\phi_3$ is approaching the vacuum on the symmetry axis, so $\phi = (0,0,1)$ as $r=0$.
Therefore there are two associated magnetic fluxes, both becomes quantized in the strong coupling limit.
More generally, the curves of $\phi_3=\pm 1$ define the structure of the magnetic fluxes of a gauged magnetic Hopfion.

\section{Initial approximation}

The task of finding of multi-soliton solutions of the Faddeev-Skyrme model
in a given sector of degree $Q$ is very complicated, it can be  performed only numerically.
Moreover, it involves rather sophisticated numerical technique see, e.g., \cite{Sutcliffe:2007ui}.

As usual, the energy minimization scheme needs an appropriate initial configuration
in a given sector. The most effective approach here is related with the
rational map approximation, suggested by Sutcliffe \cite{Sutcliffe:2007ui}.
One can consider two complex variables which parameterize the sphere $S^3$ \cite{Sutcliffe:2007ui}
\be
\left(Z_1,Z_0 \right) = \left(\sin f(r) \sin \theta e^{i\varphi};~~ \cos f(r) + i \sin f(r) \cos \theta \right) \, ,
\label{rational-Hopf}
\ee
where $f(r)$ is a monotonically decreasing function with the boundary values $f(0)=\pi$ and $f(\infty)=0$.
The coordinates $Z_1,Z_0$ are restricted to the unit sphere $S^3$, i.e.  $|Z_1|^2 + |Z_2|^2=1$. This allows us to
construct a map $\mathbb{R}^3 \mapsto S^3 \in \mathbb{C}^2$.

The components
of the scalar field $\vec\phi$, which are coordinates on the target space $S^2$,
are given by the rational map $W: S^3 \in \mathbb{C}^2 \mapsto CP^1$:
\be
W(Z_1,Z_0) = \frac{\phi_1+i\phi_2}{1+\phi_3} = \frac{P(Z_1,Z_0)}{Q(Z_1,Z_0)} \, ,
\label{rational-Hopf-2}
\ee
where the polynomials $P(Z_1,Z_0)$ and $Q(Z_1,Z_0)$  have
no common roots on the two-sphere $S^2$. The rational map ansatz \re{rational-Hopf-2}
produces a curve in $\mathbb{R}^3$, therefore
the first Hopf map $\vec \phi: \mathbb{R}^3 \mapsto S^2$ is equivalent to the rational map  $W:S^3 \mapsto CP^1$.

There are three different types of input configurations. The axially symmetric Hopfions $\mathcal{A}_{mn}$ are
produced by the rational map \cite{Sutcliffe:2007ui}
\be
W(Z_1,Z_0) =\frac{Z_1^n}{Z_0^m} \, .
\label{map-unknot}
\ee
This Hopfion has a single position curve ${\cal C}_- = \phi^{-1}(0,0,- 1)$.

More generally, we can consider initial configurations, which  are given by maps of the form
\be
W(Z_1,Z_0) =\frac{Z_1^\alpha Z_0^\beta}{Z_1^a+Z_0^b}
\label{rational-Hopf-3}
\ee
where $\alpha$ is a positive integer and $\beta$ is a non-negative integer. These maps have Hopf degree
$Q= \alpha b + \beta a$, the corresponding configuration is a torus knot $\mathcal{K}_{a,b}$.
In a particular case when $a$ and $b$ are not coprime integers, the rational map \re{rational-Hopf-3}
is degenerated producing a link  with two or more interlinked and disconnected position curves.
Configurations of that type are labeled as
$\mathcal{L}_{a,b}^{n,m}$, here the subscripts label the Hopf indexes of the unknots
and the superscript above each subscript counts the secondary linking number, which appears due to
inter-linking with the other components.

Note that the location of the soliton can be identified
as collection of curves, which follow the preimages of two distinct points, for example
$\mathcal{C}_- = \vec \phi^{-1}(0,0,-1)$ and $\mathcal{C}_{1} = \vec \phi^{-1}(1,0,0)$.
Since these loops are linked $Q$ times, the definition of the linking number can be related with
the positions of the preimages of these points: $Q= {\rm link} (\mathcal{C}_-,\mathcal{C}_{1})$. Other choice of
the preimages are also possible \cite{Samoilenka:2017hmn}.

The input for the magnetic potential $A_i$ at a finite value of the gauge coupling $g$
can be taken as a generalization of the limiting
form of the pure gauge condition above,
\be
\label{Aguess}
A_i=\pa_i\sigma(\phi_\perp) A(\phi_3) \underset{g\to\infty}{\longrightarrow} \pa_i\sigma(\phi_\perp) \, ,
\ee
where the function $\sigma(\phi_\perp)$ appeared in the
trigonometric parametrization \re{trigonometric} of the scalar fields. The smooth
function $A(\phi_3)$ must satisfy the restrictions $A(\pm1)=0$ and $A(0)\simeq1$, it agrees with the
parametrization used previously to construct axially symmetric solutions of the gauged Faddeev-Skyrme model
\cite{Shnir:2014mfa}. Thus, we can take $A(\phi_3) = 1-\phi_3^2$ as an appropriate choice.

\section{Numerical results}
For our numerical computations we used algorithm of minimization of the energy
functional \re{en} described in \cite{Samoilenka:2015bsf,Samoilenka:2016wys}.
The fields are discretized on the grid with $100^3$ or $150^3$ points with grid
spacing $\Delta x=0.1$. The initial configurations were produced via the rational map
approximation as above.
As a consistency check, we verify that our algorithm correctly reproduces the known results for the Hopfion
configurations of the usual decoupled Faddeev-Skyrme model at $g=0$ and for the
gauged axially symmetric configurations $\mathcal{A}_{11}$, $\mathcal{A}_{21}$
previously discussed in \cite{Shnir:2014mfa}.

The solutions of that type,
$\mathcal{A}_{11}$ and $\mathcal{A}_{21}$ are global minima in the sectors of degrees $Q=1,2$, respectively.
They represent axially symmetric unknots with the position curve ${\cal C}_- = \phi^{-1}(0,0,- 1)$
forming a single loop.
For the configuration $\mathcal{A}_{21}$ the corresponding linking curve, associated with preimage of the point
${\cal C}_1 = \phi^{-1}(1,0,0)$, has two twists around
the position curve, as shown in Fig.~\ref{table}.

Note that with the usual choice of the vacuum boundary conditions $\vec \phi_\infty = (0,0,1)$, the vector
$\vec n$, which appears in the definition of the covariant derivative \re{gauging}, is parallel
to $\vec \phi_\infty$. Below we will also consider another situation, when on the spacial boundary
$\vec \phi_\infty = (1,0,0)$, and $\vec n$ is transverse to $\vec \phi_\infty$.

As the gauge coupling gradually increases from $g=0$, the energy of the configuration decreases since the
magnetic flux is formed and core of the Hopfions shrinks. The magnetic energy is initially increasing,
however its contribution starts to decrease as $g$ becomes larger than $g=1$ \cite{Shnir:2014mfa}.
The structure of the magnetic field follows the pattern above,
in the weak coupling regime there is a toroidal magnetic field, which encircles the position curve of the Hopfion.
As the gauge coupling constant increases, the curves of
the preimages of the points $\mathcal{C}_{\pm} = \vec \phi^{-1}(0,0,\pm 1)$
paves the way for magnetic flux tubes.
Indeed, we can clearly identify
two fluxes along these curves, the first flux is directed alon the symmetry axis of the configuration,
the second circular magnetic flux is orthogonal to the $x-z$ plane, see Fig~\ref{table}.
Both fluxes become quantized in units of
$2\pi$ in the strong coupling limit \cite{Shnir:2014mfa}.

Further, increase of the gauge coupling leads to shrinkage of the Hopfion, magnetic field effectively squeezes
the configuration. This effect is opposite to the isorotations of the Hopfions, which also affect the structure of
the solutions \cite{Harland:2013uk,Battye:2013xf}.

\begin{figure}[t]
    \includegraphics[width=0.9\textwidth]{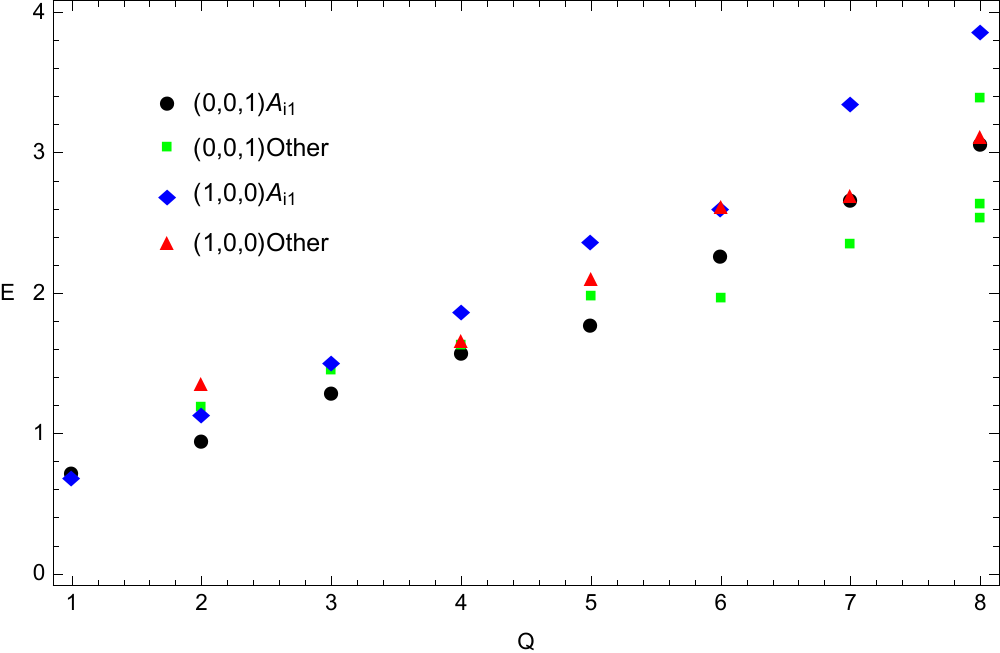}
    \caption{Energy of $Q=1-8$ Hopfions with $g=1$ and
    $\vec{\phi}_\infty=(0,0,1)$ and $\vec{\phi}_\infty=(1,0,0)$.}
    \label{EvsQ}
\end{figure}

Note that there is another  axially symmetric Hopfion configuration in the sector of degree two,
$2\mathcal{A}_{12}$ \cite{Gladikowski:1996mb,Hietarinta:1998kt,Sutcliffe:2007ui,Ward:2000qj}. It can  be thought as
two $Q=1$ Hopfions stacked one above the other. In the limit $g=0$ this solution is a saddle point
configuration, which has higher energy than
the $\mathcal{A}_{21}$ Hopfion. As $g$ increases, it still remains as a saddle point, see Fig. \ref{EvsQ}

Interestingly, there is a certain similarity between the structure of the magnetic field of
$Q=1,2$ axially symmetric Hopfions and  toroidal magnetic fields which are
well known in solar and plasma physics, see e.g. \cite{marsh}. In the latter case the magnetic field appears as a
solution of so called  force free equation for a plasma current $\vec j \times \vec B = 0$.

In order to see the difference between the cases of magnetic field of the gauged Hopfions and the magnetic field
in a plasma device, such as a stellerator or tokamak, let us assume that the scalar current $\vec j$ is a plasma current.
However, the field free equation leads to $\vec{B}=\alpha\vec{j}$,
thus for a constant $\alpha$ we obtain the Helmholtz equation:
\be\label{forcefree}
\Delta\vec{B}+\alpha^2\vec{B}=0 \, .
\ee
On the other hand, the magnetic field of the Hopfions is generated by the scalar current \re{current}, which can be
written as
\be
\vec{j} = \vec{K} (1-\phi_3^2) - {\vec \nabla} \phi_3 \times ({\vec \nabla} \phi_3 \times \vec{K}) \, ,
\ee
where $
\vec{K}={\vec \nabla} \sigma-\vec{A}$. Thus, the corresponding Maxwell equation becomes
\be
{\vec \nabla} \times({\vec \nabla}\times\vec{K}) -
2g^2 {\vec \nabla} \phi_3 \times ({\vec \nabla} \phi_3 \times \vec{K}) + 2g^2\vec{K} (1-\phi_3^2) = 0\, .
\ee
Since in the strong coupling regime the magnetic flux tubes follow the directions of the curves
of  preimages of the vectors $\vec \phi = (0,0,\pm 1)$, we can assume that outside of these curves
$\phi_3\simeq0$, thus
$$
\vec{\nabla}\times(\vec{\nabla}\times\vec{K}) + 2g^2\vec{K} \simeq 0 \, .
$$
Since $\vec{B} = \vec \nabla\times\vec{A} = -\vec \nabla\times\vec{K}$, we can see that
the counterpart of the force free equation for the magnetic field of
the Hopfions can be written as
\be
\Delta\vec{B} - 2g^2\vec{B} \simeq 0 \, .
\ee
Notably, this is a London type equation with the penetration depths parameter $\frac{1}{\sqrt2 g}$,
the mass term here has a sign opposite to the one in the force free equation \re{forcefree}.
Thus, in the strong coupling limit the magnetic field of a Hopfion
exhibit a sort of Meissner effect.

\subsection{Higher charge gauged Hopfions at $\vec{\phi}_\infty=(0,0,1)$}

Peculiar feature of the Hopfions of higher degrees is that in the standard Faddeev-Skyrme model they
usually do not possess any symmetry \cite{Sutcliffe:2007ui}, the corresponding collection of position curves
is not planar. For example, for the charge three Hopfion, the energy minimization transforms the corresponding
axially symmetric initial configuration $3\mathcal{A}_{31}$ into
the pretzel-like Hopfion $3\widetilde{\mathcal{A}}_{31}$.

\begin{figure}[h]
\includegraphics[width=1.\linewidth, angle =0]{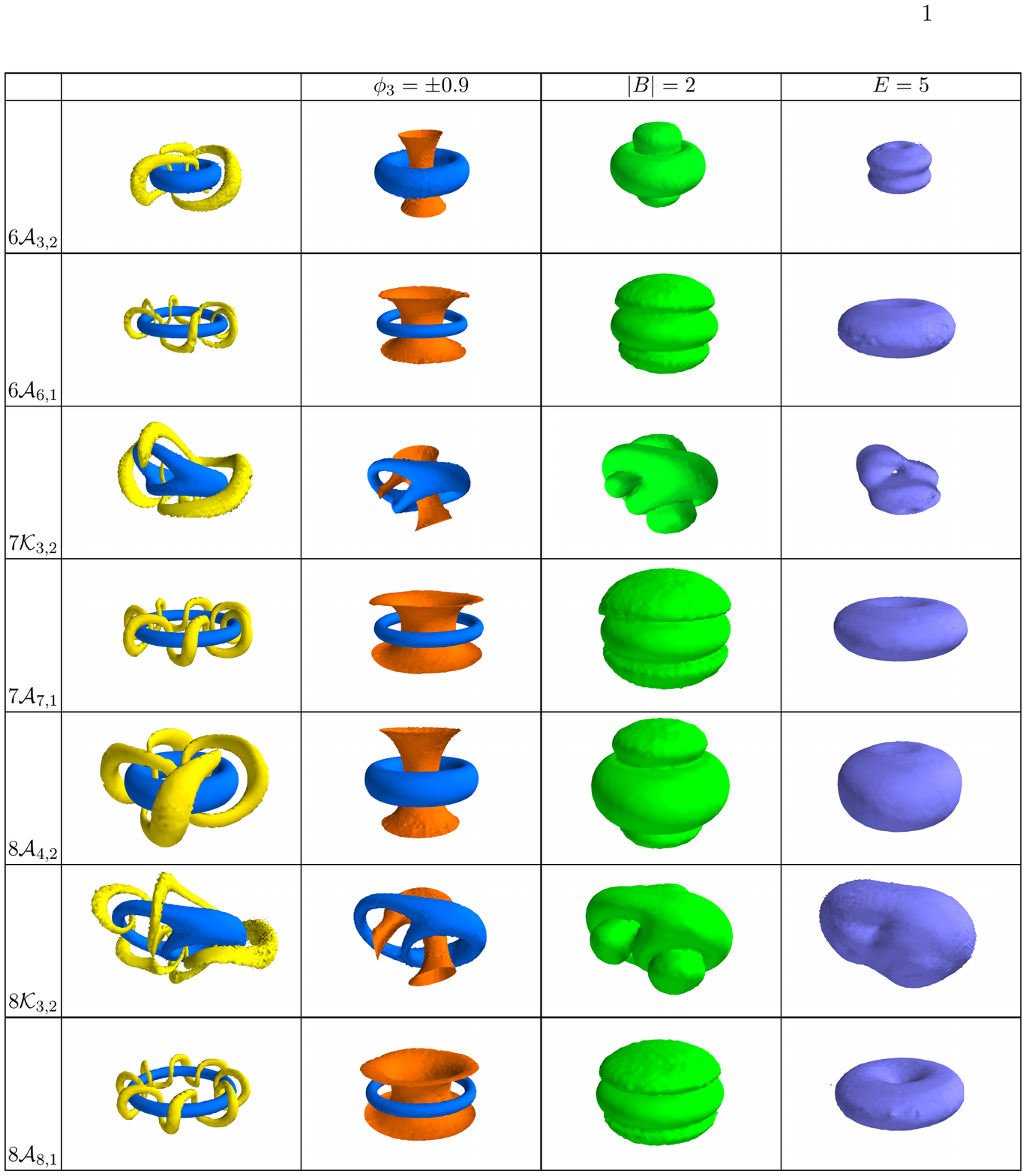}
    \caption{Isosurfaces of the field components
$\phi_1=-0.9$ and $\phi_3=-0.9$ (first column), the field components $\phi_3=\pm 0.9$ (second column),
$|B|=2$ isosurfaces of the magnetic field (third column) and $E=5$ isosurfaces of the energy density
(fourth column)  for $Q=1-8$ gauged Hopfions  in the model \re{model} with
$g=1$ and $\vec{\phi}_\infty=(0,0,1)$.}
    \label{table}
\end{figure}

Let us now consider the $Q=3$ Hopfion solution in the Faddeev-Skyrme-Maxwell model \re{model} with the usual
boundary condition $\vec{\phi}_\infty=(0,0,1)$. As the gauge coupling constant gradually increases from zero,
the position curve, initially bending toward the third direction, smoothly becomes a planar loop, see
Fig.~\ref{table}. The axially symmetric gauged Hopfion $3\mathcal{A}_{31}$ becomes the global minimum in this sector
at $g\sim 0.3$, as the gauge coupling increases further, the  deformed Hopfion
$3\widetilde{\mathcal{A}}_{31}$ does not exist as a local minimum.

Note that, similar to the case of the $Q=1,2$ Hopfions, the total energy of the configurations of higher degrees
decreases as the gauge coupling $g$ increases, this observation holds in a general case, see Fig~\ref{EvsG}.

For Hopf degree $Q=4$ there are possibilities to construct initial configurations of types
$4\mathcal{A}_{22}$, $4\mathcal{A}_{41}$ and $4\mathcal{L}_{11}^{11}$. In the usual Faddeev-Skyrme model without
the magnetic field, the axially symmetric Hopfion $\mathcal{A}_{22}$, which may be thought of as
two adjacent $2\mathcal{A}_{21}$ solitons in the maximally attractive channel of interaction,
represents the global minimum \cite{Sutcliffe:2007ui}.
Numerical relaxation of the
initial $4\mathcal{A}_{41}$ configuration yields a buckled Hopfion $4\widetilde{\mathcal{A}}_{41}$,
however in the limit $g=0$ its energy is about $2\%$ above the global minimum. Situation changes as the
gauge coupling increases, the interaction with magnetic field  tends to straighten out the position curve, thus
the axially symmetric Hopfion $4\mathcal{A}_{41}$ has lower energy than $4\CA_{22}$.

In the usual Faddeev-Skyrme model at $g=0$, in the sector of degree four
the link $4\mathcal{L}_{1,1}^{1,1}$ does not exist as a local minimum.
However, this type of solution, $5\mathcal{L}_{2,1}^{1,1}$ is a minimizer for $Q=5$ Hopfions.
We observe that increase of the gauge coupling transforms it into configuration of a different type.
As $g\sim 1$, the magnetic attraction between the fluxes, associated with the collection of loops
${\cal C}_- = \phi^{-1}(0,0,-1)$, deforms the position curve, it corresponds to two adjacent loops
which are not linked to each other, see Fig.~\ref{table}.
Further, such a configuration is not a global minimum
in this sector, as $g\gtrsim 0.2$, the axially symmetric configuration $\mathcal{A}_{51}$ has lower energy,
see Fig.~\ref{EvsG}.

Note that the magnetic field of all axially symmetric gauged Hopfions $\mathcal{A}_{n 1}$ represents
two magnetic fluxes, one flux encircles the position curve of the Hopfion
and the second one is directed along the symmetry axis, this pattern is
is illustrated in Fig.~\ref{table}. As we discussed above, in the strong coupling limit both fluxes are quantized
in units of  $2\pi$ and  $2\pi n$, respectively.

Considering the structure of the magnetic field of a $\widetilde{\mathcal{A}}_{n1}$ Hopfion  with axial symmetry
weakly broken, we observe
that one of the magnetic fluxes follows the position curve ${\cal C}_- = \phi^{-1}(0,0,- 1)$. Another flux is
associated with direction of the vector $\vec n = (0,0,1)$ in \re{gauging}. This is also the case of the $Q=5$
Hopfion, as seen in the 9th row of Fig.~\ref{table} the magnetic fluxes follow the
loops, which are preimages of $\mathcal{C}_- = \vec \phi^{-1}(0,0,-1)$ and $\mathcal{C}_{+} = \vec \phi^{-1}(0,0,1)$.

The axially symmetric configurations  $\mathcal{A}_{n 1}$ represent global minima in the strong coupling regime
up to $Q\le 5$. When $Q=6$ there are two different links $6\mathcal{L}_{2,2}^{1,1}$ and $6\mathcal{L}_{3,1}^{1,1}$,
and the axially symmetric configurations of two types, $\mathcal{A}_{3 2}$ and $\mathcal{A}_{6 1}$, respectively.
As $g=0$ the link $6\mathcal{L}_{2,2}^{1,1}$ represent the global minimum, it has a lower energy than other
configurations in that sector for all values of $g$. Increase of the coupling constant and related
magnetic interaction deforms this initial configuration into an
axially symmetric soliton $6\mathcal{A}_{3 2}$,  as $g=1$
it has energy about $16\%$ lower than another axially symmetric Hopfion $6\mathcal{A}_{6 1}$.

The trefoil knot $7\mathcal{K}_{3,2}$ is the only minimiser in the sector of degree seven as $g=0$. However, as
the gauge coupling grows, the magnetic fluxes,
associated with the loops $\mathcal{C}_- = \vec \phi^{-1}(0,0,-1)$ tends to merge because of attraction
between them. At $g=1$ the knot becomes deformed into configuration, whose position curve represents
two contacting loops, which are not interlinked,
see Fig.~\ref{table}. This structure is similar with the corresponding solution in the
sector $Q=5$.

At degree $Q=8$ there are three energy minima, which represent a link $8\mathcal{L}_{3,3}^{1,1}$, a knot
$8\mathcal{K}_{3,2}$ and axially symmetric Hopfions $8\mathcal{A}_{4 2}$, respectively. As $g=0$ the link
$\mathcal{L}_{3,3}^{1,1}$ has the energy a little less than the knot $8\mathcal{K}_{3,2}$. However, as $g$ increase,
the axial symmetry is recovered and the $8\mathcal{A}_{4,2}$ becomes global minimizer in that sector as $g=1$.
This Hopfion is composed of two $4\mathcal{A}_{41}$ Hopfions stacked one above the other with the orientation
in the maximally attractive channel. Another axially symmetric configuration $8\mathcal{A}_{8 1}$ in the strong
coupling limit has a bit higher energy. Again, we observe that the structure of the magnetic fluxes is
completely determined by the preimages of
$\mathcal{C}_- = \vec \phi^{-1}(0,0,-1)$ and $\mathcal{C}_{+} = \vec \phi^{-1}(0,0,1)$, see Fig.~\ref{table}.
As $g=1$ we find that $8\mathcal{K}_{3,2}$ Hopfion, similarly to that of $Q=7$ and $Q=5$,
deforms into configurations with two contacting loops, see Fig.~\ref{table}.

\subsection{Gauged Hopfions at $\vec{\phi}_\infty=(1,0,0)$}

Unlike gauged planar baby Skyrme model \cite{Samoilenka:2015bsf,Gladikowski:1995sc,Adam:2017ahh}, in the
3 dimensional Faddeev-Skyrme-Maxwell model \re{model} the $U(1)$ gauging prescription \re{gauging} is not
necessarily correlated with the vacuum boundary condition imposed on the scalar field.
The topological restriction on the scalar field is that  on the spacial boundary $\vec \phi$
must approach the same vacuum value regardless of direction, then $\mathbb{R}^3 \mapsto S^3$.

On the other hand, the condition of finiteness of the energy of the system for any choice of the vacuum  requires
the same restrictions on the spacial infinity
\be
D_\mu\vec{\phi}=0\, \quad F_{\mu\nu}=0 \, ,
\ee
Choosing an appropriate gauge, we can just impose $\vec{\phi}_\infty=const$.

In our consideration
above we suppose that the vector $\vec n = (0,0,1)$ and the vacuum $\vec{\phi}_\infty$ are parallel, let us
consider another possibility imposing the boundary condition
\footnote{More generally, we can consider a
continuous family of vacua
$$
\vec{\phi}_\infty=(\sin \beta\, , 0\, , \cos \beta) \, ,
$$
where parameter $\beta \in [0,2\pi]$.}
$\vec{\phi}_\infty=(1,0,0)$ with the gauging
prescription \re{gauging}. Evidently, as $g=0$ the choice of the vacuum does not affect the structure of the
Hopfion solutions in the model \re{model}  without potential,
for any particular choice of the $\vec \phi_\infty$,
the topological properties of the solitons are defined by the Hopf charge \re{charge}. However, as the gauge coupling
increases, the difference between
the directions of  $\vec \phi_\infty$ and $\vec n$ leads to significant deformations of the configurations.

\begin{figure}[h]
\includegraphics[width=1\linewidth, angle =0]{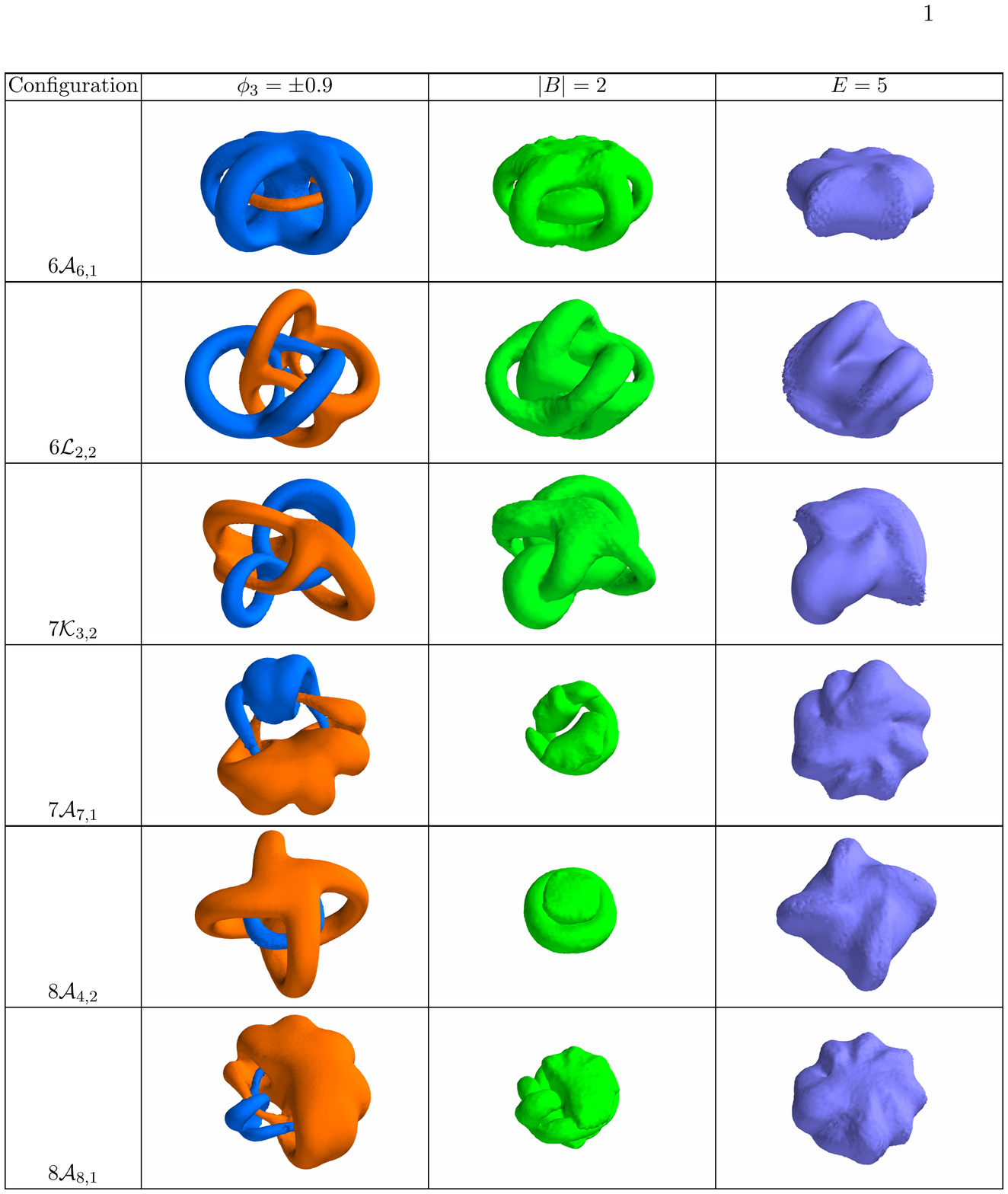}
    \caption{Isosurfaces of the field components
$\phi_1=\pm0.9$  (left column),
$|B|=2$ isosurfaces of the magnetic field (middle column) and $E=5$ isosurfaces of the energy density
(right column)  for $Q=1-8$ gauged Hopfions  in the model \re{model} with
 with $g=1$ and $\vec{\phi}_\infty=(1,0,0)$.}
    \label{table_rot}
\end{figure}

The results of energy minimization simulations are summarized in Fig.~\ref{table_rot}.
We confirm that the structure of the magnetic field of the gauged Hopfion is always determined by
the collection of loops ${\cal C}_\pm = \phi^{-1}(0,0,\pm 1)$, for any value of the parameter $\beta$.

We observe that as the parameter $\beta$ is increasing, the energy of the $Q=1$ Hopfion is decreasing, however,
the static energy of the configurations of higher degrees is increasing. Rotation of the vacuum with respect to the
direction of the vector $\vec n$ effectively deforms the soliton, as $g=1$ and $\vec{\phi}_\infty=(1,0,0)$,
the $Q=1$ Hopfion is no longer axially symmetric, the magnetic field represent two fluxes, which
are linked once. Although the position curve of this Hopfion remains a single loop, the energy density distribution
of the configuration at $g\sim 1$ looks more like the link, see Fig.~\ref{table_rot}.

Slightly deformed at $g=1$  unknot $2\mathcal{A}_{21}$ is the minimal energy configuration in the sector $Q=2$,
see Figs.~\ref{table_rot}, \ref{EvsG}. Interestingly, as the gauge coupling increases,
the position curve of the higher energy solution $2\widetilde{\mathcal{A}}_{12}$ is splitting into two
contiguous loops, see Fig.~\ref{table_rot}, third row.
However, unlike the position curve of a link, there is no
inter-linking of two loops. Thus, this configuration can be labeled as ${\mathcal{L}}_{1,1}^{0,0}$.
The magnetic fluxes, which correspond to the curve ${\cal C}_+ = \phi^{-1}(0,0,1)$, are propagating in the
same direction.

More generally, the splitting
loops are touching each other without interlinking at the point where magnetic fluxes are parallel.
Our calculations show that it may happened
at the center of the Hopfion, like for
$2\CA_{2,1},\ 3\CA_{3,1},\ 4\CA_{2,2},\ 4\CA_{4,1},\ 6\CA_{6,1},\ 8\CA_{4,2}$ configurations,
see Fig.~\ref{table_rot}. The point  of contact can also be not at the center of the configuration,
it happens for $2\CA_{1,2},\ 5\CL_{1,2},\ 6\CA_{2,2}$ Hopfions.

Another possibility is that the curves of ${\cal C}_\pm = \phi^{-1}(0,0,\pm 1)$ lie on top of each other, we
observed this type of behavior for $3\CA_{3,1},\ 4\CA_{2,2},\ 4\CA_{4,1},\ 5\CL_{1,2}$ Hopfions.

Considering the axially symmetric gauged Hopfions  $\CA_{i,j}$ we found that
they may either form symmetric configurations with a single loop ${\cal C}_- = \phi^{-1}(0,0,-1)$, like
$2\CA_{2,1}, 4\CA_{4,1}, 6\CA_{6,1}$ etc, or the blobs may appear on the loops, like
$5\CA_{5,1},\ 7\CA_{7,1},\ 8\CA_{8,1}$ configurations, see Fig.~\ref{table_rot}. Notably, the
$6\CA_{6,1}$ and $8\CA_{4,2}$ Hopfions possess $D_6$ symmetry and $D_4$ symmetries, respectively.

\begin{figure}[h]
\includegraphics[width=1.\linewidth, angle =0]{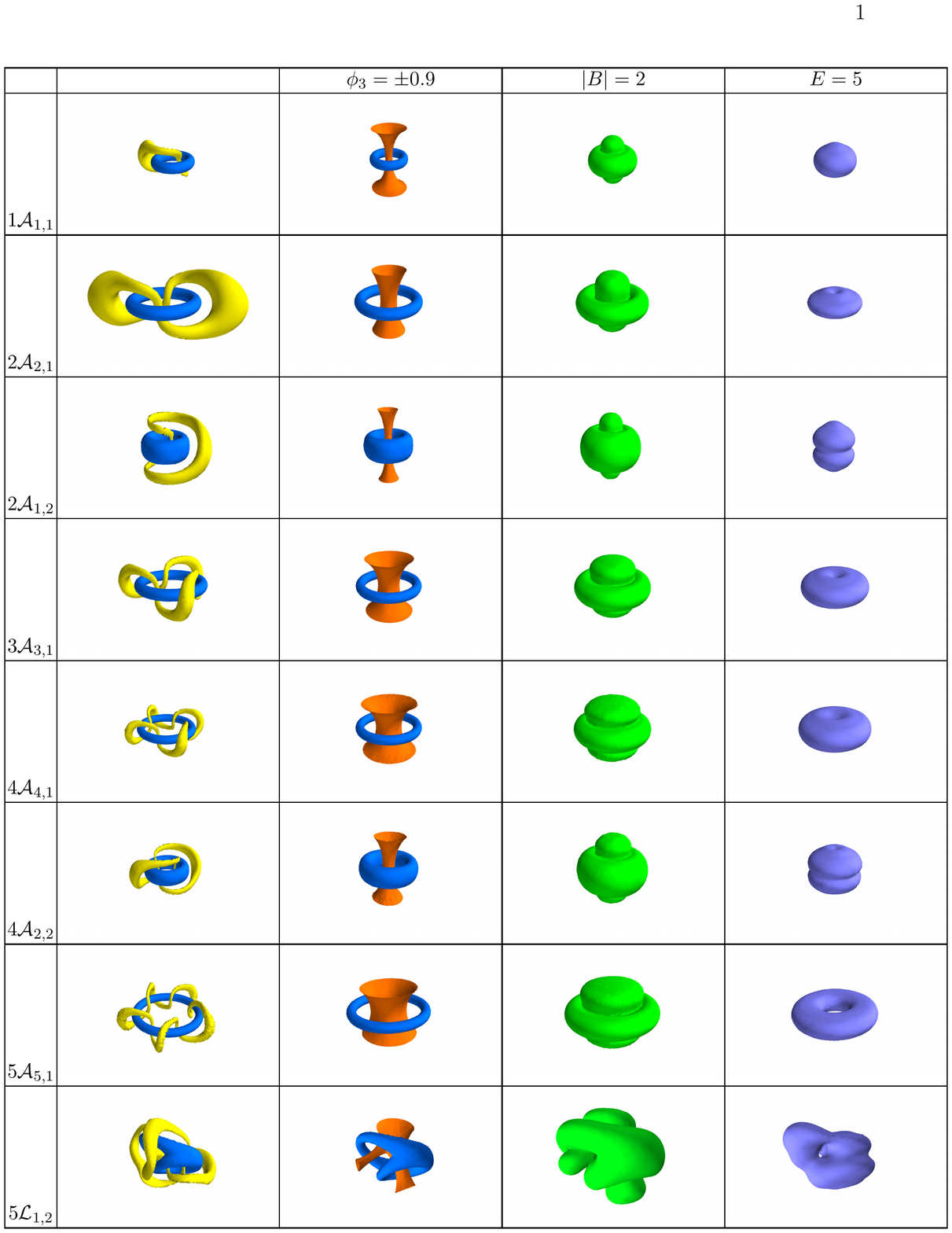}
\end{figure}

\begin{figure}[h]
\includegraphics[width=1\linewidth, angle =0]{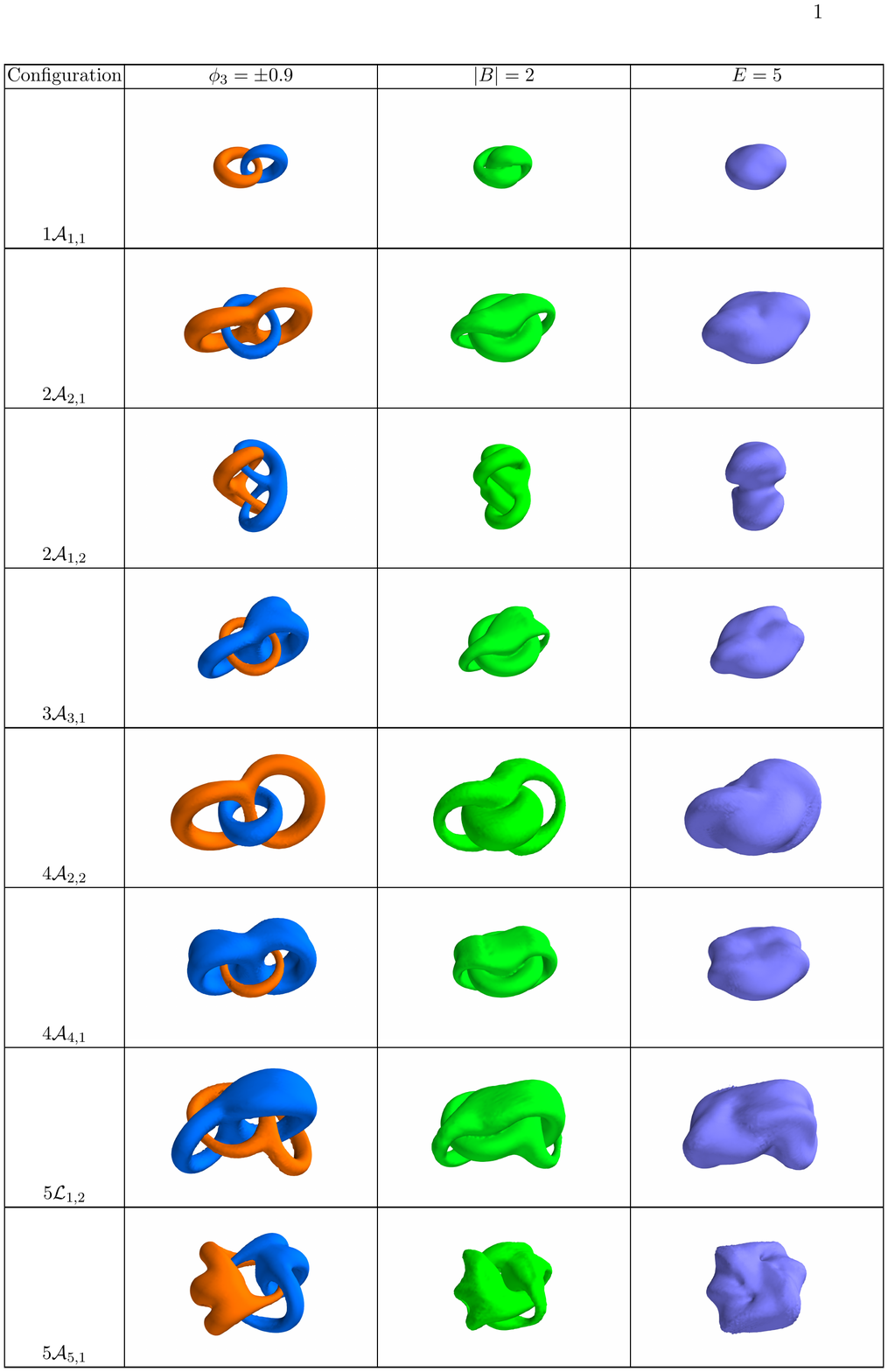}
\end{figure}

\subsection{Dependence on gauge coupling for $Q=5$}
As a particular example of the parametric dependency of the gauged Hopfions on the coupling constant $g$, we
considered solitons in the sector of degree $Q=5$, both in the case of the vacuum
$\vec{\phi}_\infty=(0,0,1)$ and $\vec{\phi}_\infty=(1,0,0)$.

\begin{figure}[h]
    \centering
    \includegraphics[width=.49\textwidth]{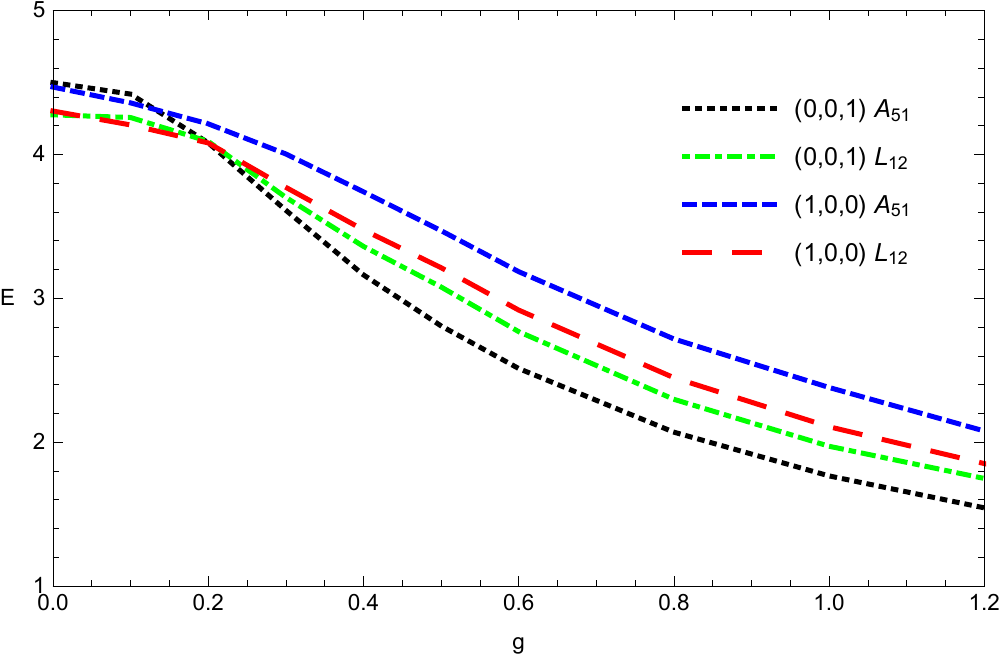}
    \includegraphics[width=.49\textwidth]{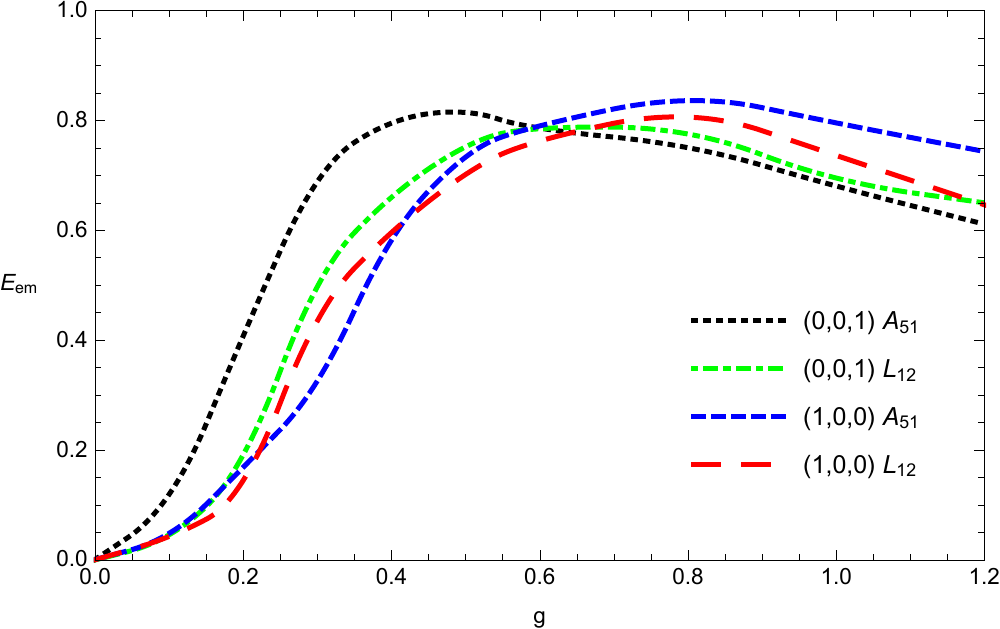}
    \caption{ The normalized energy $E$ of the $5\CA_{5,1}$ and $5\mathcal{L}_{1,2}^{1,1}$ gauged Hopfions
(left plot) and the corresponding magnetic energy (right plot)
as a function of the coupling constant $g$ in the Faddeev-Skyrme-Maxwell model \re{model} for
the vacua $\vec{\phi}_\infty=(0,0,1)$ and $\vec{\phi}_\infty=(1,0,0)$.}
\label{EvsG}
\end{figure}

In Fig.~\ref{EvsG} we have plotted the graphs of total energy of the
gauged $5\CA_{5,1}$ and $5\CL_{1,2}^{1,1}$ Hopfions,
defined by the functional \re{en}, and magnetic energy as function of the gauge
coupling $g$. As $g=0$ both choices of the vacuum are equivalent,
the energy of the link $5\mathcal{L}_{1,2}^{1,1}$ is lower than the axially symmetric Hopfion $5\CA_{5,1}$.
However, as the gauge coupling increases from zero, the latter configuration becomes a
global minimum in the vacuum $\vec{\phi}_\infty=(0,0,1)$, while the link $5\CL_{1,2}^{1,1}$ still remains
the minimal energy solution in the vacuum $\vec{\phi}_\infty=(1,0,0)$,
as seen in the left plot, Fig.~\ref{EvsG}.

General observation is that, as the gauge coupling
increases, the energy of the gauged Hopfion monotonically decreases. On the other hand,
the magnetic energy initially increases
from zero, it attains its maximum  at $g\simeq0.7$. Further increase of the coupling leads to decrease of the magnetic
energy, as shown in the right plot, Fig.~\ref{EvsG}. As expected, the size of the Hopfions decreases as the coupling
$g$ increases.

\begin{figure}[h!]
    \centering
\includegraphics[width=0.9\linewidth, angle =0]{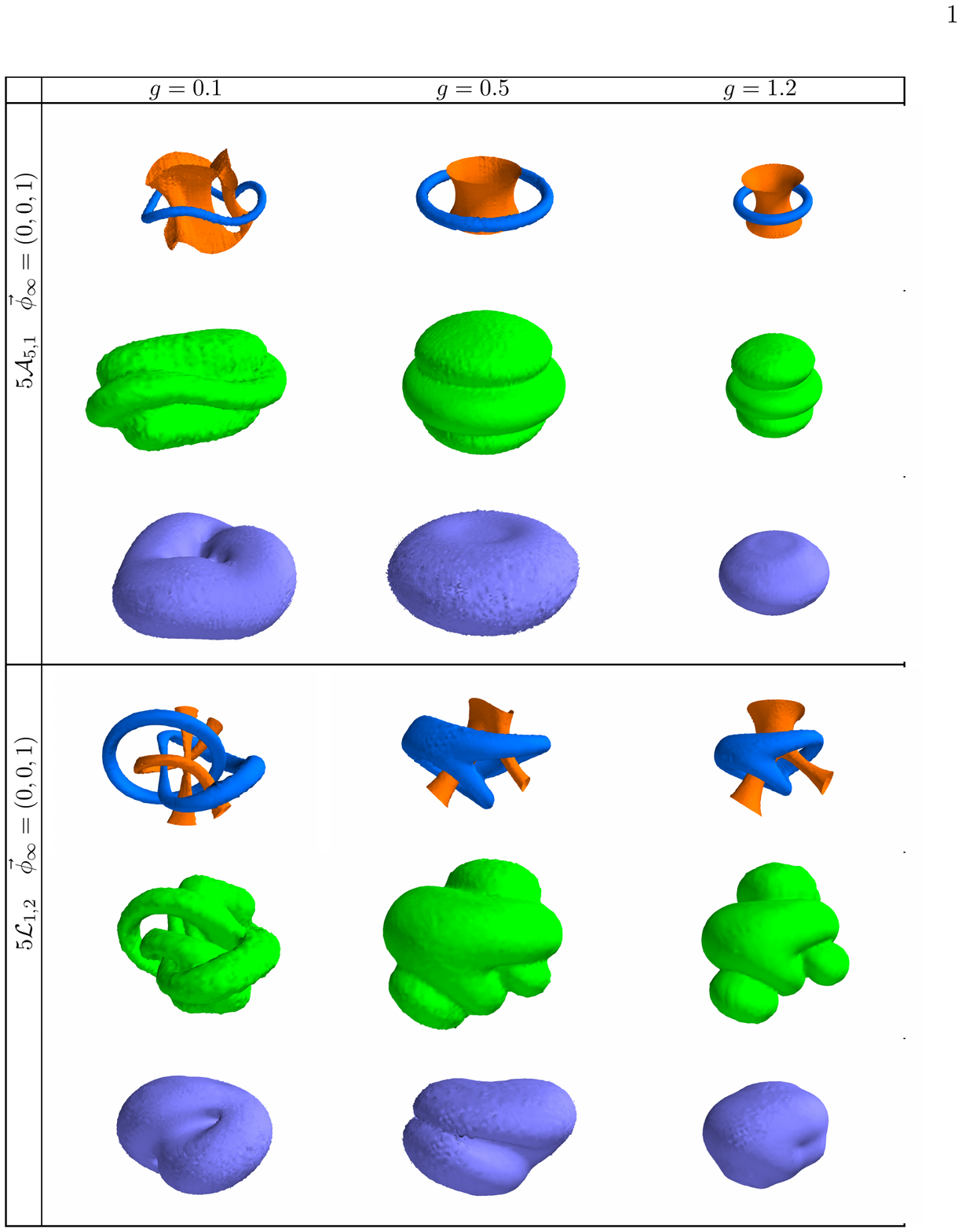}
    \caption{Isosurfaces of the field components $\phi_3=\pm0.9$ (first row),
$|B|=1$ isosurfaces of the magnetic field (second row)
and  $E=2$ isosurfaces of the energy density (third row) of the $Q=5$ Hopfions in the model \re{en} with the vacuum
 $\vec{\phi}_\infty=(0,0,1)$ for
$g=0.1\, \ g=0.1 $ and $g=1.2$.}
\label{table_q5}
\end{figure}

\begin{figure}[h]
    \centering
\includegraphics[width=0.9\linewidth, angle =0]{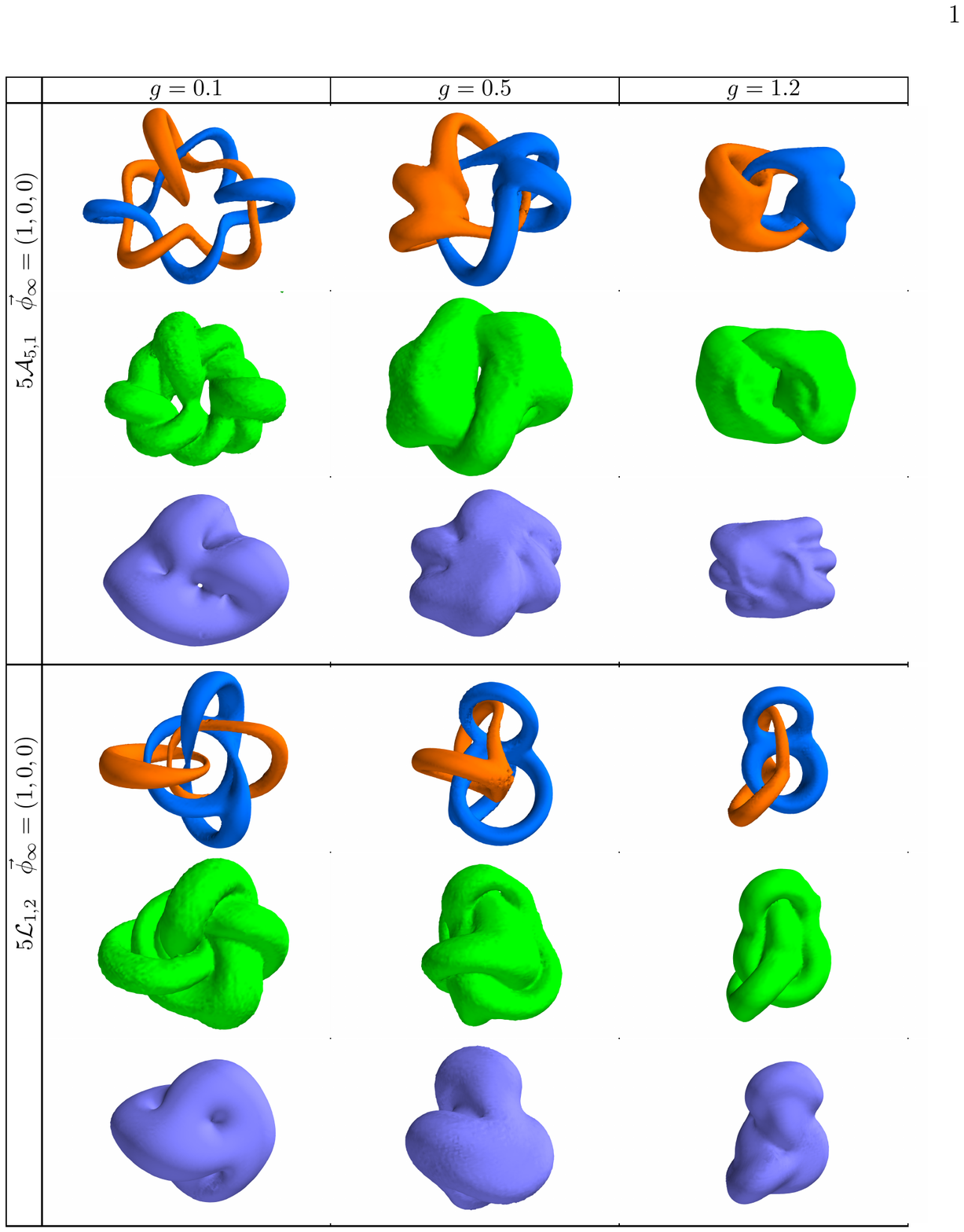}
    \caption{Isosurfaces of the field components $\phi_3=\pm0.9$ (first row),
$|B|=1$ isosurfaces of the magnetic field (second row)
and  $E=2$ isosurfaces of the energy density (third row) of the $Q=5$ Hopfions in the model \re{en}
with the vacuum $\vec{\phi}_\infty=(1,0,0)$ for
$g=0.1\, \ g=0.1 $ and $g=1.2$.}
\label{table_q5_90}
\end{figure}

In Figs.~\ref{table_q5},\ref{table_q5_90} we display the pattern of evolution of the initial
$5\CA_{5,1}$ and $5\CL_{1,2}^{1,1}$ configurations, as the gauge coupling $g$ is growing from zero.
In the model \re{model} with the usual choice of the vacuum $\vec{\phi}_\infty=(0,0,1)$, coupling to the magnetic
field, directed along the vector $\vec n \parallel \vec{\phi}_\infty$, recovers the axial symmetry of the
$5\CA_{5,1}$ Hopfion, which is violated as $g \lesssim 0.25$. Within that range of values of $g$ the
bent axial solution $5\widetilde \CA_{5,1}$ is a local energy minimum,
the minimal energy configuration in this sector remains the link $5\CL_{1,2}^{1,1}$. However, as $g \gtrsim 0.25$
the lowest energy solution is the axially symmetric Hopfion $5\CA_{5,1}$. The magnetic fluxes
of this configuration are directed through the center of the Hopfion and around the symmetry axis, as shown in
Fig.~\ref{table_q5}, upper panel.
In the strong coupling limit the link $5\CL_{1,2}^{1,1}$ becomes strongly deformed, see Fig.~\ref{table_q5}, bottom
panel. Magnetic attraction between the fluxes, associated with position curve of the soliton,
deforms the curve itself, for sufficiently large values of $g$ it is shaped as
two contacting loops without interlinking. As we have seen above,
the magnetic fluxes and the energy density distribution follow the curves ${\cal C}_\pm = \phi^{-1}(0,0,\pm 1)$.

The pattern of the evolution of the $Q=5$ Hopfions in the \re{model} with the
the vacuum $\vec{\phi}_\infty=(1,0,0)$, following the increase of $g$, is somewhat
different from what is outlined above. The position curve of the $Q=5$ axially symmetric configuration
$\CA_{5,1}$ gradually becomes deformed into a loop with internal twisting, see Fig.~\ref{table_q5_90}, upper
panel. The link $5\CL_{1,2}^{1,1}$ has lower energy as $g \gtrsim 0.4$, the position curve of this configuration,
is deformed into two twisted unlinked adjoining rings.

\section{Conclusions}

The objective of this work is to investigate properties of soliton solutions of the Faddeev-Skyrme-Maxwell
model. We have considered the Hopfion solutions with topological charges up to $Q=8$,
coupled to the magnetic field. We found that, as the gauge coupling increased, the backreaction of the magnetic field
may significantly affect the structure of the solutions, however, the results
depend both on the type of the vacuum boundary condition and
on the strength of the gauge coupling.
We found that the
magnetic fluxes of gauged Hopfions follow the directions provided by preimages of the vectors
$\vec \phi = (0,0,\pm 1)$. In the strong coupling limit the magnetic field of the gauged Hopfion
exhibits behavior similar to the field of the vortex solution of the abelian Higgs model.

The work here should be taken further by considering the electrically charged configurations, another
interesting direction is to investigate the soliton solutions of the $SO(3)$ gauged Faddeev-Skyrme model.
It might be also interesting to consider gauged Hopfions
in frustrated magnets, which combine nearest-neighbour ferromagnetic and
higher-neighbour  anti-ferromagnetic interactions \cite{Sutcliffe:2017aro}.
We hope we can address these issues in our future work.

\section*{Acknowledgements}
Y.S. gratefully
acknowledges support from the Ministry of Education and Science
of Russian Federation, project No 3.1386.2017.
The parallel computations were performed on the cluster HIBRILIT at LIT, JINR, Dubna.

\begin{small}

\end{small}


\begin{thebibliography}{13}
\bibitem{BB}
A.A.~Bogolubskaya  and I.L.~Bogolubsky,
Phys.\ Lett.\ A  {\bf 136} (1989) 485\\
A.A.~Bogolubskaya  and I.L.~Bogolubsky,
Lett.\ Math.\ Phys.\  {\bf 19} (1990)  171.
\bibitem{Leese:1989gj}
R.A.~Leese, M.~Peyrard and W.J.~Zakrzewski,
Nonlinearity {\bf 3} (1990) 773.
\bibitem{Leese:1989gj}
R.A.~Leese, M.~Peyrard and W.J.~Zakrzewski,
Nonlinearity {\bf 3} (1990) 773.
\bibitem{Skyrme:1961vq}
 T.~H.~R.~Skyrme,
  Proc.\ Roy.\ Soc.\ Lond.\ A   {\bf 260} (1961) 127.
\bibitem{Adam:2010fg}
  C.~Adam, J.~Sanchez-Guillen and A.~Wereszczynski,
  Phys.\ Lett.\ B {\bf 691} (2010) 105
\bibitem{Adam:2010ds}
  C.~Adam, J.~Sanchez-Guillen and A.~Wereszczynski,
  Phys.\ Rev.\ D {\bf 82} (2010) 085015
\bibitem{Faddeev-Hopf}L.D.~Faddeev, {\it{"Quantization of Solitons,"}},
Preprint-75-0570, IAS, Princeton (1975)
\bibitem{Faddeev:1996zj}L.D.~Faddeev and A.J.~Niemi, Nature {\bf 387} (1997) 58
\bibitem{Derrick}G.H. Derrick, J.\ Math.\ Phys.\  {\bf 5}, (1964) 1252
\bibitem{Kobayashi:2013xoa}M.~Kobayashi and M.~Nitta, Phys.\ Lett.\ B {\bf 728} (2014) 314
\bibitem{Foster:2010zb}D.~Foster,  Phys.\ Rev.\ D  {\bf 83} (2011) 085026
\bibitem{Sutcliffe:2011ig}P.~Sutcliffe,
  JHEP {\bf 1104} (2011) 045
\bibitem{Aratyn:1999cf}
  H.~Aratyn, L.~A.~Ferreira and A.~H.~Zimerman,
  Phys.\ Rev.\ Lett.\  {\bf 83} (1999) 1723
\bibitem{Ferreira:2017yzy}
  L.~A.~Ferreira and Y.~Shnir,
  Phys.\ Lett.\ B {\bf 772} (2017) 621
\bibitem{Ferreira:2017bsr}
  L.~A.~Ferreira,
  JHEP {\bf 1707} (2017) 039
\bibitem{Gladikowski:1996mb}
  J.~Gladikowski and M.~Hellmund,
  Phys.\ Rev.\  D {\bf 56} (1997) 5194
\bibitem{Battye1998}
R.~Battye and P.~Sutcliffe,
Phys.\ Rev.\ Lett.\  {\bf 81}  (1998) 4798
\bibitem{Sutcliffe:2007ui}
P.~Sutcliffe,
Proc.\ Roy.\ Soc.\ Lond.\  A {\bf 463} (2007) 3001
\bibitem{Manton-book}N.~Manton and P.~Sutcliffe, {\it Topological Solitons},
(Cambridge University  Press, Cambridge, 2004)
\bibitem{Shnir-book}Y.M.~Shnir, \textit{Topological and Non-Topological Solitons in Scalar Field Theories},
(Cambridge University Press, Cambridge, 2018)
\bibitem{Battye:2004rw}R.A.~Battye and P.M.~Sutcliffe: Nucl.\ Phys.\ B {\bf 705} (2005) 384
\bibitem{Battye:2009ad}R.A.~Battye, N.S.~Manton, P.~Sutcliffe and S.W.~Wood: Phys.\ Rev.\ C {\bf 80} (2009) 034323
\bibitem{Salmi:2014hsa}P.~Salmi and P.~Sutcliffe, J.\ Phys.\ A  {\bf 48}  (2015) 035401
\bibitem{Gudnason:2016cdo}
  S.~B.~Gudnason and M.~Nitta,
  Phys.\ Rev.\ D {\bf 94} (2016) no.6,  065018
\bibitem{Gudnason:2016yix}
  S.~B.~Gudnason and M.~Nitta,
  Phys.\ Rev.\ D {\bf 94} (2016) no.2,  025008
\bibitem{jaykka2010easy}
  J.~Jaykka and M.~Speight,
  Phys.\ Rev.\ D {\bf 82} (2010) 125030
\bibitem{kobayashi2013fractional}
M.~Kobayashi and M.~Nitta,
  J.\ Low.\ Temp.\ Phys.\  {\bf 175} (2014) 208
\bibitem{kobayashi2014vortex}
M.~Kobayashi and M.~Nitta,
  Phys.\ Rev.\ D {\bf 87} (2013) no.12,  125013
\bibitem{Gudnason:2016tiz}
  S.~B.~Gudnason, B.~Zhang and N.~Ma,
  Phys.\ Rev.\ D {\bf 94} (2016) no.12,  125004
\bibitem{Gudnason:2015nxa}
  S.~B.~Gudnason and M.~Nitta,
  Phys.\ Rev.\ D {\bf 91} (2015) no.8,  085040
\bibitem{Lukacs:2016mvy}
  \'{A}.~Luk\'{a}cs,
  J.\ Math.\ Phys.\  {\bf 57} (2016) no.7,  072903
\bibitem{Samoilenka:2017hmn}
  A.~Samoilenka and Y.~Shnir,
  JHEP {\bf 1709} (2017) 029
\bibitem{Gladikowski:1995sc}
  J.~Gladikowski, B.~M.~A.~G.~Piette and B.~J.~Schroers,
  Phys.\ Rev.\ D {\bf 53} (1996) 844
\bibitem{Samoilenka:2015bsf}
A.~Samoilenka and Y.~Shnir,
Phys.\ Rev.\ D {\bf 93} (2016)  065018
\bibitem{Shnir:2014mfa}
  Y.~Shnir and G.~Zhilin,
  Phys.\ Rev.\ D {\bf 89} (2014) no.10,  105010
\bibitem{Harland:2013uk} D.~Harland, J.~J\"akk\"a, Y.~Shnir and M.~Speight,
J.\ Phys.\ A
{\bf 46} (2013) 225402
\bibitem{Battye:2013xf}RA.~Battye and M.~Haberichter,
Phys.\ Rev.\ D {\bf 87} (2013)   105003
\bibitem{Hietarinta:1998kt}J.~Hietarinta and P.~Salo,
Phys.\ Lett.\ B  {\bf 451}   (1999) 60
\bibitem{Ward:2000qj}R.S.~Ward,
Phys.\ Lett.\ B {\bf 473}   (2000) 291
\bibitem{Schroers:1995he}
B.J.~Schroers,
Phys.\ Lett.\ B  {\bf 356} (1995) 291.
\bibitem{Gladikowski:1995sc}
J.~Gladikowski, B.~M.~A.~G.~Piette  and  B.~J.~Schroers,
Phys.\ Rev.\ D  {\bf 53} (1996) 844.
\bibitem{Adam:2017ahh}
  C.~Adam and A.~Wereszczynski,
  Phys.\ Rev.\ D {\bf 95} (2017) no.11,  116006
\bibitem{Samoilenka:2017fwj}
  A.~Samoilenka and Y.~Shnir,
  Phys.\ Rev.\ D {\bf 97} (2018) no.4,  045004
\bibitem{Shnir:2009ct}
  Y.~Shnir and D.~H.~Tchrakian,
  J.\ Phys.\ A {\bf 43} (2010) 025401
\bibitem{Samoilenka:2016wys}
  A.~Samoilenka and Y.~Shnir,
  Phys.\ Rev.\ D {\bf 95} (2017) no.4,  045002
 \bibitem{marsh}
  Gerald E. Marsh, {\it Force-Free Magnetic Fields: Solutions, Topology and Applications}, World Scientific (1996).
\bibitem{Sutcliffe:2017aro}P.~Sutcliffe,
Phys.\ Rev.\ Lett.\  {\bf 118} (2017)  247203.

\end{thebibliography}

\end{document}